\documentclass[useAMS,usenatbib]{mn2e}

\usepackage{graphics,graphicx}
\usepackage{paralist}
\usepackage{amsmath}
\usepackage{amssymb}
\usepackage{lscape}
\usepackage{rotating}
\usepackage{afterpage}
\usepackage{color}
\usepackage{multirow}

\title[Total X-ray emission from star-forming galaxies]{X-ray emission from star-forming galaxies - III. Calibration of the {\boldmath $L_{\rmn{X}}-\rmn{SFR}$} relation up to redshift {\boldmath  $z\approx 1.3$}}

\author[S. Mineo, M. Gilfanov, B. D. Lehmer,  G. E. Morrison, and R. Sunyaev]{S. Mineo$^{1,2,3}$\thanks{E-mail: smineo@head.cfa.harvard.edu}, M. Gilfanov $^{2, 4}$, B. D. Lehmer$^{5,6}$, G. E. Morrison$^{7,8}$ and R. Sunyaev $^{2, 4}$\\
$^{1}$Harvard-Smithsonian Center for Astrophysics, 60 Garden Street Cambridge, MA 02138, USA\\
$^{2}$Max Planck Institut f\"ur Astrophysik, Karl-Schwarzschild-Str. 1 85741 Garching, Germany\\
$^{3}$Department of Physics, University of Durham, South Road, Durham DH1 3LE, UK\\
$^{4}$Space Research Institute of Russian Academy of Sciences, Profsoyuznaya 84/32, 117997 Moscow, Russia\\
$^{5}$The Johns Hopkins University, Homewood Campus, Baltimore, MD 21218, USA\\
$^{6}$NASA Goddard Space Flight Centre, Code 662, Greenbelt, MD 20771, USA\\
$^{7}$Institute for Astronomy, University of Hawaii, Manoa, HI 96822, USA\\
$^{8}$Canada France Hawaii Telescope Corp., Kamuela, HI 96743, USA
}
\begin{document}
\sloppypar

\date{Accepted. Received; in original form}

\pagerange{\pageref{firstpage}--\pageref{lastpage}} \pubyear{2010}

\maketitle

\label{firstpage}

\begin{abstract}

We investigate the relation between total X-ray emission from star-forming
galaxies and their star formation activity. Using nearby late-type galaxies and
ULIRGs from Paper~I and star-forming galaxies from {\it Chandra} Deep Fields,
we construct a sample of 66 galaxies  spanning the redshift range $z\approx
0-1.3$ and the star-formation rate (SFR) range $\sim
0.1-10^3\,M_{\odot}\,\rmn{yr}^{-1}$. In agreement with previous results, we
find that the $L_{\rmn{X}}-\rmn{SFR}$ relation is consistent with a linear law
both at $z=0$ and for the $z=0.1-1.3$ CDF  galaxies, within the statistical
accuracy of $\sim 0.1$ in the slope of the $L_{\rmn{X}}-\rmn{SFR}$ relation.
For the total sample, we find a linear scaling relation $L_{\rmn{X}}/\rmn{SFR}\approx (4.0\pm 0.4) \times
10^{39}(\rmn{erg}\,\rmn{s}^{-1})/(M_{\odot}\,\rmn{yr}^{-1})$, with a scatter
of $\approx 0.4$ dex. About $\sim 2/3$ of the 0.5--8 keV luminosity generated
per unit SFR is expected to be due to HMXBs. We find no statistically
significant trends in the mean $L_{\rmn{X}}/\rmn{SFR}$ ratio with the redshift
or star formation rate and constrain the amplitude of its variations by
$\lesssim 0.1-0.2$ dex. These properties make X-ray observations a powerful
tool to measure the star formation rate in normal star-forming galaxies that
dominate the source counts at faint fluxes.   

\end{abstract}


\section{Introduction}

High-mass X-ray binaries (HMXBs) and the hot ionized inter-stellar medium (ISM)
are the main contributors to the total X-ray output of normal (i.e. not
containing a luminous active galactic nucleus (AGN)) star-forming galaxies.  It
is well established that the collective X-ray luminosity of HMXBs well
correlates with the star formation activity of the host galaxy
\citep[][hereafter Paper~I]{2003MNRAS.339..793G, 2003A&A...399...39R,
2010ApJ...724..559L, 2011AN....332..349M, 2012MNRAS.419.2095M}. The  hot
ionized ISM contributes about $\sim 1/4$ to the observed X-ray emission from
late-type galaxies in the standard X-ray band (0.5--8 keV), its
luminosity has been also shown to scale linearly with star formation rate (SFR)
\citep[][hereafter Paper~II]{2005ApJ...628..187G, 2012arXiv1210.2997L,
2012MNRAS.426.1870M}. Thus, it has been proposed that the total, integrated
X-ray luminosity from star-forming galaxies can be used as a proxy of the SFR
\citep{2003MNRAS.339..793G, 2003A&A...399...39R}. Although not entirely free
from its own systematic uncertainties and contaminations, the X-ray based SFR
proxy  is less affected by the interstellar extinction and cosmological
passband redshift, than conventional SFR indicators. Furthermore, the
$L_{\rmn{X}}-\rmn{SFR}$ scaling relation does not experience significant
cosmological evolution up to redshifts of $z\sim 1-2$. This has been initially
suggested based on direct measurements of the $L_{\rmn{X}}/\rmn{SFR}$ ratios
for several galaxies in Chandra Deep Fields \citep{2003MNRAS.339..793G,
2008ApJ...681.1163L, 2010ApJ...724..559L, 2012MNRAS.419.2095M} and was further  supported by
calculations of the maximal contribution of X-ray faint star-forming galaxies
to the unresolved part of the Cosmic X-ray background
\citep{2012MNRAS.421..213D} and stacking analysis results
\citep{2012ApJ...748...50C}. These properties  make the X-ray based SFR proxy a
powerful tool to measure the star formation rate in distant galaxies. 

The most significant systematic effect which can compromise the X-ray-based SFR
measurements is the contamination by the emission of the central supermassive
black hole, the AGN. Indeed, even low luminosity AGN, with
$\log(L_{\rmn{X}})\sim 42$ can outshine a $\sim 100 M_\odot$/yr starburst. As
populations of bright galaxies are mainly composed of AGN \citep[for recent
results, see e.g.][]{2011ApJS..195...10X, 2012ApJ...752...46L}, SFR
measurements using X-ray luminosity can be applied  only to a relatively small
fraction of bright galaxies. In this case, a careful investigation of the
nature of each galaxy is required in order to separate late-type from
early-type galaxy populations. On the contrary, among faint sources,
$F_{\rmn{X}}\la 10^{-17}$ erg~cm$^{-2}$~s$^{-1}$, the majority are star-forming galaxies
located at moderate and large redshifts, $z\sim 0.5-3$
\citep{2012ApJ...748...50C,2012ApJ...752...46L}.  This makes the X-ray based
SFR proxy a powerful tool to measure the star formation rate in faint galaxies,
where it can be used {\em en masse}, to infer the cosmic star formation history
\citep[e.g.][]{2012ApJ...748...50C}.

The aim of this paper is to obtain the  $L_{\rmn{X}}-\rmn{SFR}$  scaling
relation for the {\em total} X-ray luminosity and to investigate its behavior
in a broad range of redshifts.  For the redshift $z=0$, we use  the  sample of
star-forming galaxies and ULIRGs (ultra-luminous infrared galaxies) from Papers
I and II. We then select normal star-forming galaxies from the {\it Chandra}
Deep Fields (CDFs)  expanding  the local sample towards cosmologically
interesting redshifts and high star formation rates. We combine these data in
order to calibrate the $L_{\rmn{X}}-\rmn{SFR}$ scaling relation over a broad
range of redshifts and star formation rates.

The structure of the paper is as follows. In Section 2 we briefly summarize the
selection criteria and properties of  the local sample. In Section 3 we
describe selection of late-type galaxies from the CDF
data and the procedures used to calculate their X-ray luminosities and star
formation rates. The $L_{\rmn{X}}-\rmn{SFR}$  relation is derived in Section 4
and its redshift and SFR dependences are investigated in Section 5.  In Section 6 we
summarize our findings.

Throughout this paper we assume a flat $\Lambda$CDM cosmology with $H_0 = 70$
km/s/Mpc, $\Omega_M = 0.3$ and $\Omega_\Lambda = 0.7$.

\section{Local galaxies}
\label{sec:resolved_sample}

\begin{table*}
\centering
\begin{minipage}{120mm}
\caption{The local ($z=0$) sample of star-forming galaxies.}
\label{table:local_sample}
\begin{tabular}{@{}l c c c c c c @{}}
\hline
 Galaxy & D & Ref.\footnote{References for distances: (1) \citet{2003AJ....126.1607S}, (2) \citet{1988Sci...242..310T}, (3) \citet{2008A&A...486..151G}, (4) \citet{2003Ap.....46..144M}, (5) \citet{1997ApJS..109..333W}, (6) \citet{2006ApJS..165..108S}, (7) \citet{2005JKAS...38....7I}, (8) \citet{2004AJ....127..660S}, (9) \citet{2002AJ....124..811D}, (10) \citet{2000A&AS..142..425D}, (11) \citet{2009AJ....138..323T}.}  & Hubble type & SFR\footnote{Star formation rate from Spitzer and GALEX data (see Sect. 6 Paper~I).} &  $M_{\star}$\footnote{Stellar mass from 2MASS data (see Sect. 5 Paper~I).}  & $\log L^{\rmn{tot}}_{\rmn{X}}$\footnote{Total X-ray luminosity of galaxies, i.e. gas and X-ray binary contribution, in 0.5--8 keV band, measured as described in Sect. \ref{sec:nearby_Xray_lums}.}\\
 & (Mpc) & & &  ($M_{\odot}\,\rmn{yr}^{-1}$) & ($10^{10}\,M_{\odot}$) & $(\rmn{erg}\,\rmn{s}^{-1})$ \\ 
\hline
\hline
\multicolumn{7}{|c|}{{\sc resolved galaxies}}\\
\hline
NGC~0278	& $11.8^{+2.4}_{-2.0}$ & (2) & SAB(rs)b & 4.1  & 0.7 & 39.52  \\ 
NGC~0520 	& $27.8^{+5.6}_{-4.7}$  & (2) & Pec & 11.6 & 4.7  & 40.53  \\ 
NGC~1313 	& $4.1\pm0.2$ & (3) & SB(s)d & 0.44 & 0.1 & 39.75  \\
NGC~1569 	& $1.9\pm 0.2$ & (4) & IB  & $7.8\times 10^{-2}$ & $2.8\times 10^{-2}$ & 38.17  \\ 
NGC~2139 	& $26.7^{+5.8}_{-4.8}$ & (5) & SAB(rs)cd & 3.8  & 0.91 & 40.48  \\ 
NGC~3079 	& 18.2 & (1) & SB(s)c & 6.0  & 4.0  & 40.30  \\ 
NGC~3310 	& 19.8 & (1) & SAB(r)bc pec & 7.1  & 0.98  & 41.02  \\ 
NGC~3556 	& $10.7^{+1.9}_{-1.6}$ & (11) & SB(s)cd  & 3.1  & 1.7  & 39.87  \\ 
NGC~3631 	& $24.3\pm 1.6$ & (7) & SA(s)c & 4.6  & 2.6  & 40.79 \\ 
NGC~4038/39 	& $13.8\pm 1.7$  & (8) & - & 5.4  & 3.1  & 40.57 \\ 
NGC~4194 	& $39.1^{+7.9}_{-6.6}$ & (2) & IBm pec & 16.8 & 2.1 & 40.92  \\ 
NGC~4214 	& $2.5\pm 0.3$ & (9) & IAB(s)m & 0.17 & $3.3\times 10^{-2}$  & 38.53  \\ 
NGC~4490 	& $7.8^{+1.6}_{-1.3}$& (2) & SB(s)d pec & 1.8  & 0.39 & 40.32  \\ 
NGC~4625 	& $8.2^{+1.7}_{-1.4}$ & (2) & SAB(rs)m pec & 0.09 & $7.0\times 10^{-2}$  & 37.92 \\ 
NGC~5253 	& $4.1\pm 0.5$ & (6) & Im pec & 0.38 & $5.9\times 10^{-2}$  & 38.31  \\ 
NGC~5474 	& 6.8 & (10) & SA(s)cd pec & 0.18 & $9.1\times 10^{-2}$  & 39.05  \\ 
NGC~5775 	& $26.7^{+11.9}_{-8.2}$ & (2) & Sb(f) & 5.3  & 6.3  & 40.88  \\ 
NGC~7090 	& 7.6 & (1) & SBc & 0.29 & 0.22  & 39.08  \\ 
NGC~7541 	& $34.9^{+6.6}_{-5.7}$ & (5) & SB(rs)bc pec & 14.7 & 4.9  & 40.39  \\ 
NGC~7793 	& $4.0^{+0.7}_{-0.6}$ & (11) & SA(s)d & 0.29 & 0.15  & 38.53   \\ 
UGC~05720 	& $24.9^{+11.1}_{-7.7}$ & (2) & Im pec & 1.8  & 0.37  & 39.77  \\ 
\hline
\multicolumn{7}{|c|}{{\sc unresolved galaxies}}\\
\hline
IRAS~17208-0014  &183.0 & (1) & ULIRG & 289.9 & 10.3  & 41.40  \\
IRAS~20551-4250  & 179.1 & (1) & ULIRG & 139.4 & 7.5   & 41.63   \\
IRAS~23128-5919  & 184.2 & (1) & ULIRG & 139.6 & 7.0   & 41.88   \\
IRAS~10565+2448 & 182.6 & (1) & ULIRG & 156.8 & 9.9  & 41.42   \\
IRAS~13362+4831 & 120.9 & (1) & LIRG & 54.8  & 17.3   & 41.81 \\
IRAS~09320+6134 & 164.3 & (1) & ULIRG & 137.1 & 13.1  & 41.39  \\
IRAS~00344-3349  & 84.0  & (1) & LIRG & 20.1  & 1.5   & 41.23   \\
NGC~4676	        & 98.2  &  (1) & - & 15.8  & 6.6   &  40.92  \\
\hline
\end{tabular}
\end{minipage}
\end{table*}

\subsection{The sample}

We constructed the local sample of star-forming galaxies by merging a sample of
nearby galaxies resolved by {\it Chandra} from Paper~II (Sect. 2, Table 1), and
a sample of unresolved LIRGs (luminous infrared galaxies) and ULIRGs defined in
Paper~I (Sect. 2.2, Table 2). The  local sample contains 29 star-forming
galaxies, their parameters are summarized in Table \ref{table:local_sample}.

The SFRs and stellar masses ($M_{\star}$)  were determined in Paper~I,  based
on far-infrared, UV and K-band luminosities. These quantities were measured for
the same spatial regions as used for computing the X-ray luminosities, defined
in Section 5 of Paper~I. The star formation rates and stellar masses of the
local sample span  broad ranges, from  $\sim 0.1-
290\,M_{\odot}\,\rmn{yr}^{-1}$ and $\sim 3\times 10^{8}- 2\times
10^{11}\,M_{\odot}$, respectively.

\begin{figure}
\centering
{
\includegraphics[trim=1mm 15mm 5mm 5mm, width=1.0\linewidth]{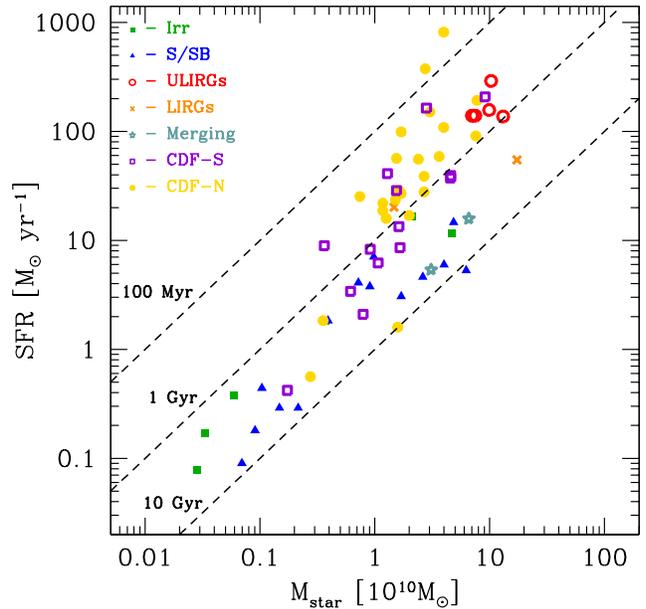}
}
\caption{The $\rmn{SFR}-M_{\star}$ plane. Different types of galaxies are
plotted with different symbols: Irregulars in (green) filled squares, spirals
in (blue) filled triangles, ULIRGs in (red) empty circles, LIRGs in (orange)
crosses, interacting systems in (cyan) empty stars. Star-forming galaxies
from CDF-S and CDF-N are plotted as empty squares
(purple) and filled circles (gold) respectively. The dashed lines correspond to
constant stellar-mass-to-SFR ratio.  }
\label{fig:m-sfr}
\end{figure}

\subsection{Total X-ray luminosity}
\label{sec:nearby_Xray_lums}
 
The data preparation was done following standard
CIAO\footnote{http://cxc.harvard.edu/ciao3.4/index.html} threads (CIAO version
3.4; CALDB version 3.4.1), exactly the same way as described in Sect. 3 of
Paper~I.  
 
For resolved galaxies, listed in the first part of Table
\ref{table:local_sample}, source spectra were extracted in the 0.5--8 keV
energy band and background spectra were created as described in Sect. 3.3 of
Paper~II. Spatial regions for extracting source spectra were defined in
Paper~I.  Both source and background spectra and the associated ARF and RMF
files were produced by using \texttt{specextract} script and modeled using
XSPEC v.  12.3.1x.  The spectra were grouped in order to have a minimum of 15
counts per channel to apply the $\chi^2$ fitting. For most of the spectra a
good fit was obtained with a two component model: a thermal plasma ({\tt
mekal}) plus a power-law, corrected for the Galactic absorption.  A second
photoelectric absorption component, accounting for the intrinsic absorption
and/or a second thermal component, was required for some of the galaxies.  The
addition of such a component was decided on a galaxy-by-galaxy basis via 
the
result of the  F-test, using a probability of $10^{-3}$ as the threshold
(see Paper~II for details). For each galaxy, we used the best fit model to
calculate  the count-to-erg conversion coefficient. Using this coefficient, the
energy flux was computed form the background subtracted source count rate.  

For the sample of unresolved starburst galaxies presented in the second part of
Table \ref{table:local_sample}, we adopted 0.5--8 keV luminosities from
Paper~I.

\section{Star-forming galaxies from Chandra Deep Fields}
\label{sec:cdf_sel}

\begin{table*}
\centering
\begin{minipage}{160mm}
\caption{Late-type galaxies from the \textit{Chandra} Deep Fields.}
\label{table:cdf_galaxies}
\begin{tabular}{@{}l c c c c c c c c@{}}
\hline
CXO & Redshift & Field\footnote{References for redshift and fields: E-CDF-S and CDF-N galaxies are from \citet{2008ApJ...681.1163L}, 4Ms CDF-S galaxies are from \citet{2011ApJS..195...10X}.} & Morph.\footnote{Morphological Type from \citet{2005ApJ...625..621B}, according to the following classification: 3=Sab, 4=S, 6=Irr, 8=Merger. Sources without numerical morphological type in column 4 were classified by eye based on HST images as described in Sect.\ref{sec:type} for details.}  & $\log M_{\star}$ & $\log L_{0.5-8 \rmn{keV}}$ & $S_{1.4 \rmn{GHz}}$ & $\log L_{1.4 \rmn{GHz}}$ & SFR \\
 & & & Type & ($M_{\odot}$) & (erg s$^{-1}$) & ( $\umu$Jy) & (erg s$^{-1}$) & ($M_{\odot}$ yr$^{-1}$) \\
\hline
J033209.80-274442.9  &	0.08  & 4Ms CDF-S  & 3 & 9.24	& 39.57	& 45.2	 &  27.88 & 0.4 \\
J033211.54-274713.3  &	0.58  & 4Ms CDF-S  & 6 & 10.45 & 41.08	& 239.3	 &  30.47 & 163.9 \\
J033217.90-275100.1  &	0.12  & 4Ms CDF-S  & 4 & 9.90	& 39.52	& 108.9	 &  28.58 & 2.1 \\
J033220.28-275222.3  &	0.34  & 4Ms CDF-S  & 6 & 10.21 & 40.69	& 64.5	 &  29.38 & 13.4 \\
J033223.67-274938.5  &	0.58  & 4Ms CDF-S  & 6 & 10.19 & 40.91	& 41.5	 &  29.71 & 28.7 \\
J033227.73-275041.2  &	1.10  & 4Ms CDF-S  & 6 & 10.96 & 41.67	& 66.0	 &  30.57 & 207.7 \\
J033228.00-274639.7  &	0.25  & 4Ms CDF-S  & 3 & 10.22 & 40.01	& 86.3	 &  29.19 & 8.6 \\
J033229.86-274425.1  &	0.08  & 4Ms CDF-S  & 4 & 9.96	& 39.94	& 1068.0 &  29.17 & 8.3 \\
J033230.00-274405.0  &	0.07  & 4Ms CDF-S  & 3 & 9.79	& 40.09	& 452.2	 &  28.79 & 3.4 \\
J033232.99-275030.0  &	0.67  & 4Ms CDF-S  & 4 & 10.66 & 41.15	& 38.2	 &  29.83 & 37.4 \\
J033236.19-274932.0  &	0.55  & 4Ms CDF-S  & 8 & 10.66 & 41.31	& 64.6	 &  29.85 & 39.2 \\
J033237.27-275127.5  &	0.69  & 4Ms CDF-S  & 6 & 10.11 & 41.23	& 38.9	 &  29.87 & 40.9 \\
J033238.83-274956.5  &	0.25  & 4Ms CDF-S  & 3 & 10.03 & 40.04	& 63.9	 &  29.05 & 6.2 \\
J033256.66-275316.5  & 	0.37  & E-CDF-S  & 8 & 9.56 & 41.39	& 36.3	 &  29.20 & 8.9 \\
J123559.74+621550.3  & 0.38 & CDF-N & 4 &     10.56 & 41.28 &  225.8 & 30.02 &  58.8  \\ 
J123619.47+621252.9  & 0.47 & CDF-N & 3 &     10.43 & 41.31 &  65.3  & 29.70 &  27.8  \\
J123621.09+621208.5  & 0.78 & CDF-N & 6 &     10.43 & 41.45 &  27.6  & 29.84 &  38.7  \\
J123623.03+621346.7  & 0.48 & CDF-N & 8 &     10.07 & 41.39 &  41.8  & 29.53 &  18.7  \\
J123634.47+621213.4  & 0.46 & CDF-N & 8 &     10.88 & 41.53 &  224.7 & 30.21 &  91.0  \\
J123634.53+621241.3  & 1.22 & CDF-N & 6 &     10.6  & 42.39 &  201.1 & 31.17 &  813.9 \\
J123636.86+621135.2  & 0.08 & CDF-N & 4 &     9.44  & 39.27 &  65.0  & 28.00 &  0.6   \\
J123639.93+621250.4  & 0.85 & CDF-N & 4 &     10.6  & 41.87 &  63.3  & 30.29 &  108.8 \\
J123643.97+621250.4  & 0.56 & CDF-N & 4 &     10.18 & 41.07 &  36.8  & 29.63 &  23.6  \\
J123645.40+621901.3  & 0.45 & CDF-N & 3 &     10.1  & 41.4  &  41.3  & 29.46 &  15.9  \\
J123646.60+621049.4  & 0.94 & CDF-N & -&     10.19 & 41.62 &  25.9  & 30.01 &  56.5  \\
J123646.67+620833.6  & 0.97 & CDF-N & 3 &     10.89 & 41.93 &  81.7  & 30.54 &  192.0 \\
J123648.30+621426.9  & 0.14 & CDF-N & 3 &     9.55  & 40.05 &  64.7  & 28.52 &  1.8   \\
J123649.72+621313.4  & 0.47 & CDF-N & 6 &     9.87  & 41.14 &  59.3  & 29.66 &  25.6  \\
J123651.12+621031.2  & 0.41 & CDF-N & 4 &     10.23 & 41.1  &  86.8  & 29.69 &  26.9  \\
J123652.77+621354.7  & 1.36 & CDF-N & -&     10.48 & 42.15 &  28.8  & 30.43 &  150.6 \\
J123652.83+621808.1  & 0.25 & CDF-N & 3 &     10.3  & 40.53 &  167.4 & 29.48 &  16.9  \\
J123653.37+621140.0  & 1.27 & CDF-N & 8 &     10.44 & 42.3  &  84.2  & 30.83 &  374.7 \\
J123658.34+620958.4  & 0.14 & CDF-N & 3 &     10.2  & 39.98 &  56.4  & 28.46 &  1.6   \\
J123659.92+621450.3  & 0.76 & CDF-N & 8 &     10.38 & 41.45 &  42.1  & 30.00 &  55.5  \\
J123700.41+621617.3  & 0.91 & CDF-N & -&     10.23 & 41.76 &  49.0  & 30.25 &  99.0  \\
J123708.32+621056.4  & 0.42 & CDF-N & 6 &     10.07 & 41.26 &  66.8  & 29.60 &  21.9  \\
J123716.82+621007.8  & 0.41 & CDF-N & 3 &     10.21 & 40.89 &  92.7  & 29.71 &  28.8  \\
\hline
\end{tabular}
\end{minipage}
\end{table*}
\label{sec:cdf}

In order to extend the $L_{\rmn{X}}-\rmn{SFR}$ relation towards higher
redshifts and SFRs, we selected a sample of late-type galaxies in the Chandra
Deep Fields. We started from the samples studied by
\citet{2008ApJ...681.1163L}, who selected late-type galaxies in the 2 Ms
Chandra Deep Field-North \citep[CDF-N;][]{2003AJ....126..539A} and Extended Chandra Deep Field-South
\citep[E-CDF-S;][]{2005ApJS..161...21L}, and by \citet{2011ApJS..195...10X}, the 4 Ms Chandra Deep
Field-South (CDF-S) survey.  For the E-CDF-S, the
\citet{2008ApJ...681.1163L} late-type galaxy sample was constructed using data
within a 1 Ms {\it Chandra} exposure of the CDF-S proper
\citep{2003AJ....126..539A} and four flanking E-CDF-S fields with 250 ksec {\it
Chandra} exposures each \citep{2005ApJS..161...21L}.  The
\citet{2011ApJS..195...10X} 4~Ms exposure covered only the CDF-S but to
$\approx$4--16 times larger depth. 

As we describe in the sections below, to study the $L_{\rm X}$--SFR correlation and
its redshift dependence, we filtered these samples according to the following
criteria: (i) rigorous determination of the morphological type appropriate for
a star-forming galaxy; (ii) availability of a robust 0.5--8 keV X-ray
detection; (iii) availability of 1.4~GHz radio flux measurement for the SFR
calculation or SFR determinations by \citet{2008ApJ...681.1163L}; (iv)  the
value of specific SFR ($\mathrm{SFR}/M_{\star}$) larger than
$10^{-10}\,\rmn{yr}^{-1}$.

\subsection{Selection by morphological type}
\label{sec:type}

We began by applying two main filtering criteria to the
\citet{2008ApJ...681.1163L} and \citet{2011ApJS..195...10X} CDF
catalogs: \begin{inparaenum}[\itshape i\upshape)] we excluded \item all
the AGN candidates and \item objects detected only in one of the sub-bands
(i.e., 0.5--2~keV and 2--8~keV), for which full band luminosity was not
available. \end{inparaenum} Then we proceeded with more accurate filtering, excluding 
the early type galaxies using the morphological classifications of galaxies in 
the GOODS fields \citep{2005ApJ...625..621B} and original {\em HST} data for those 
galaxies which were not included in the GOODS catalogs.

The 4 Ms CDF-S survey \citep{2011ApJS..195...10X}
includes 740 X-ray sources, of which 578 are classified as AGN and 162 as
normal galaxies.  The additional filtering mentioned above reduced the sample to
102 normal CDF-S galaxies.  We cross-correlated these sources with the morphological
catalog of galaxies in the GOODS South field \citep{2005ApJ...625..621B}. Using
a matching radius of $1.5\arcsec$, we found 79 matches. Among them, we selected
only sources classified as late-type or merging galaxies. 
The 22 sources with no match in the \citet{2005ApJ...625..621B}
catalog have  $z_{850}$-band magnitude fainter than the
threshold limit applied to construct the above mentioned catalog. We
constructed {\it HST} $B_{435}$, $V_{606}$, and $z_{850}$ color image cut-outs
and classified each galaxy as either early-type or late-type by visual
classification. Seven sources were classified as early type
galaxies and consequently were excluded from further analysis. We also excluded one 
source because its morphology did not allow us to clearly classify it neither as a late-type nor 
as an early-type galaxy. The remaining 14 sources were added to the selection of CDF-S 
sources obtained so far. After our morphological selection, we were left with a sample of 67
star-forming galaxies from the 4 Ms CDF-S region.

The sample from \citet{2008ApJ...681.1163L} includes 225 X-ray detected
late-type galaxies (selected via optical colors).  Among them, 121 were
classified as AGN candidates, while the remaining 104 sources were classified as
normal late-type galaxies. Applying the same morphological filtering criteria
as above yielded 52 sources: 37 in the CDF-N and 15 in the E-CDF-S.  As
\citet{2012ApJ...752...46L} point out, a single color division does not
perfectly disentangle the early-type and late-type galaxy populations. In order
to exclude a possible early-type galaxy contamination of the
\citet{2008ApJ...681.1163L} sample, we cross-matched the 52 sources with the
morphological catalogs of galaxies in the GOODS North and South fields
\citep{2005ApJ...625..621B}. Using a matching radius of $1.5\arcsec$ we found a
total of 44 matches, 11 for the E-CDF-S and 33 for the CDF-N sample. Two
galaxies in the latter sample (J123603.26+621111.3 and J123627.32+621258.1) were
classified as early-type and were excluded.  We inspected the {\it HST} color image
cut-outs of the four sources with no match in the GOODS-North morphological
catalog.  We classified three of these sources as late-type galaxies based on visual inspection. Concerning the E-CDF-S late-type galaxies from
\citet{2008ApJ...681.1163L}, 10 of these sources were already part of the
\citet{2011ApJS..195...10X} sample.  Hereafter, we adopt the properties from the deeper 4 Ms CDF-S from
\citet{2011ApJS..195...10X} for these galaxies.
The obtained sample from \citet{2008ApJ...681.1163L} consists of 35 unique
star-forming galaxies.

As a check, we cross-correlated the obtained sample of CDF-N sources with the
``High-SFR sample'' in Paper~I (Table 3). This sample was compiled by
\citet{2003MNRAS.339..793G} and was used in Paper~I as a secondary sample of
unresolved star-forming galaxies in order to explore the
$L_{\rmn{X}}-\rmn{SFR}$ relation in the high SFR regime. It includes seven
galaxies from {\it Chandra} observations of the {\it Hubble Deep Field North}
\citep[HDF-N;][]{2001AJ....122.2810B} and one from the Lynx field
\citep{2002AJ....123.2223S}.  For the seven HDF-N sources, we found six
counterparts. The remaining galaxy, 123716.3+621512, was not present in the
GOODS catalog. We constructed the {\it HST} image cut-out as 
described above and visually classified the source as an early-type galaxy. 
Correspondingly, we did not consider it further in our analyses.

After this selection we were left with a sample of 104 X-ray selected
star-forming galaxies drawn from the CDFs.  The sample will be
reduced further after the cross-correlation with radio catalogs (see Sect.
\ref{sec:cdf_sfr}).

\subsection{X-ray luminosity}
\label{sec:cdf_Xray_lums}

Both \citet{2008ApJ...681.1163L} and \citet{2011ApJS..195...10X} provide
rest-frame X-ray luminosities in the 0.5--8 keV band. In both cases the
K-correction was performed assuming a power spectrum.
\citet{2008ApJ...681.1163L} used a photon index $\Gamma=2$, appropriate for
star-forming galaxies \citep[e.g.][and references
therein]{2005AJ....129....1L}, whereas \citet{2011ApJS..195...10X} used a
photon index of $\Gamma=1.8$, more appropriate for AGNs
\citep[e.g.][]{2006A&A...451..457T}. For sources from the
\citet{2008ApJ...681.1163L} sample we used their original luminosity values. However,
recomputed luminosities for galaxies from \citet{2011ApJS..195...10X}, using
their fluxes and redshifts and adopting a photon index
$\Gamma=2$.

\subsection{Cross-correlation with radio catalogs and star formation rate estimates}
\label{sec:cdf_sfr}

The far-infrared (FIR) and ultraviolet (UV) luminosities are good proxies
of the star formation activity in late-type galaxies. However, the use of these
two estimators for CDF galaxies is limited by both the lack of rest-frame UV
spectroscopic data and the sensitivity limit of the available IR observations.
For these reasons, \citet{2008ApJ...681.1163L} were able only to compute upper
limits on SFRs for most of the sources in their galaxy sample. To measure star
formation rates in the CDF galaxies, we used 1.4~GHz radio emission
\citep[e.g.][]{1992ARA&A..30..575C}. It has been shown
\citep[][]{2003ApJ...586..794B, 2006ApJ...643..173S} that this estimator
is in good agreement with both IR- and UV-based estimators used for the local
resolved galaxy sample (Sect. \ref{sec:resolved_sample}).  We cross-correlated
the X-ray sample of star-forming galaxies produced in section \ref{sec:type}
with catalogs of radio sources based on VLA observations of CDFs. For the
CDF-N we used the \citet{2010ApJS..188..178M} catalog, and Miller et al. (2013) catalog for the CDF-S. Using a matching radius of $1.5\arcsec $ we
found a total of 39 matches with VLA sources: 23 from CDF-N and 16 from the
CDF-S, of which one uniquely belongs to the selection from
\citet{2008ApJ...681.1163L} and 15 to the 4 Ms CDF-S sample of
\citet{2011ApJS..195...10X}. The radio flux densities $S_{1.4\,\rmn{GHz}}$ were
converted to luminosities using the following equation: 

\begin{equation} 
\label{eq:lr_vla} 
L_{1.4\,\rmn{GHz}} = 4 \pi d^{2}_{L}
S_{1.4\,\rmn{GHz}} (1+z)^{\alpha-1}. 
\end{equation}

For CDF-N galaxies
detected in the \citet{2010ApJS..188..178M} catalog, we used the total (i.e.
integrated) radio flux densities $S_{1.4\,\rmn{GHz}}$. For CDF-N galaxies in
the Miller et al. (2013) catalog, we used either peak or integrated
flux densities according to their prescription. 
We assumed a spectral index of $\alpha = 0.8$, according 
to \citet{1992ARA&A..30..575C}. The radio-based SFRs were
estimated using the calibration of \citet{2003ApJ...586..794B}:
\begin{equation} 
\label{eq:sfr_radio} 
\rmn{SFR}\,(M_{\odot}\,\rmn{ yr}^{-1}) =
5.55\times 10^{-29}L_{1.4\,\rmn{GHz}} (\rmn{erg}\,\rmn{s}^{-1}) 
\end{equation}

\subsection{X-ray sources with ``missing'' radio counterparts}
\label{sec:cdf_sfr_ul}

As discussed in Section \ref{sec:cdf_sfr},  only a fraction X-ray selected
normal late-type galaxies have radio counterparts: 23 out of 35 CDF-N sources
($\sim 66\%$) and 16 out of 68 CDF-S sources ($\sim 24\%$).

\begin{figure}
\centering
{
\includegraphics[trim=1mm 20mm 5mm 5mm, width=1.0\linewidth]{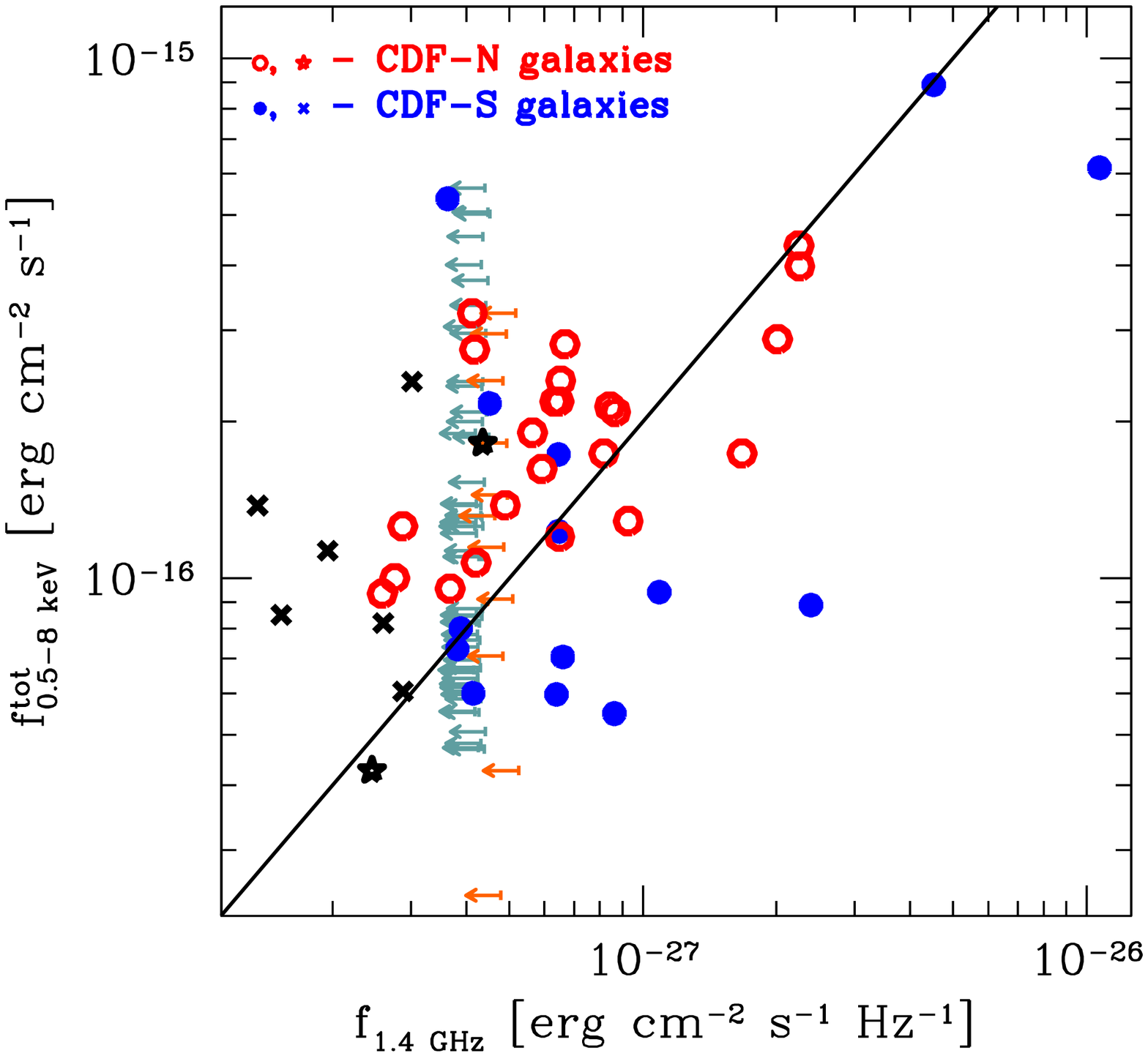}
}
\caption{1.4~GHz flux versus 0.5--8 keV flux for the whole sample of CDFs. The
plot shows 39 late-type galaxies in both North (filled blue circles) and South
(empty red circles) fields, that were detected in both X-ray
\citep{2008ApJ...681.1163L, 2011ApJS..195...10X} and radio \citep[][Miller et
al. 2013]{2010ApJS..188..178M} (see Sect.  \ref{sec:cdf_sel}). Black empty stars and crosses indicate sources 
with radio detection significance between $3\sigma<\rm{SNR}<5\sigma$ in CDF-N and CDF-S respectively. 
These sources were not included in the radio catalogs mentioned above (Sect.~\ref{sec:cdf_sfr_ul}). The 64 
X-ray detected late-type galaxies with radio upper limits are plotted with cyan (CDF-S) and orange (CDF-N) 
arrows at $5\sigma$. The solid line is our best fitting $L_{\rmn{X}}-\rmn{SFR}$
relation, eq. (\ref{eq:ltot_sfr_all}), converted in flux units.} 
\label{fig:fx_fr}
\end{figure}

In Fig.~\ref{fig:fx_fr} we plot the 1.4~GHz flux versus 0.5--8 keV flux for the
entire sample of CDF sources classified as normal late-type galaxies. For
sources undetected in radio we plot upper limits. The latter were derived from
the publicly-available radio RMS maps of both the GOODS-N VLA Deep 20cm
Radio Survey\footnote{http://www.ifa.hawaii.edu/$\sim$morrison/GOODSN/}
\citep{2010ApJS..188..178M} and the VLA 1.4GHz Survey of the Extended Chandra
Deep Field South: Second Data
Release\footnote{http://www.astro.umd.edu/$\sim$nmiller/VLA\_ECDFS.html}
(Miller et al. 2013).  The plotted upper limits in Fig.~\ref{fig:fx_fr} are at
the $5\sigma$ level.  For reference, the solid line shows our best fitting
$L_{\rmn{X}}-\rmn{SFR}$ relation derived in Section \ref{sec:lx_sfr_relation},
eq.  (\ref{eq:ltot_sfr_all}), converted to flux units. To calculate the solid line, the average 
value of redshift among CDF galaxies is assumed. It can be seen in Fig.~\ref{fig:fx_fr} that the
radio-undetected sources span both X-ray bright as well as X-ray faint sources
with little variation in the radio limits.  If the galaxies with the highest
X-ray fluxes, at the few$\times 10^{-16}\,\rm{erg}\,\rm{cm}^{-2}\,\rm{s}^{-1}$ level, followed the
average $L_X-$SFR relation, we would have expected them to be detected in the
radio band. On the other hand, some the X-ray faint sources may be normal star-forming 
galaxies that have remained undetected because of the insufficient sensitivity of the radio data.

In order to investigate the nature of the X-ray sources undetected in the radio
band, we stacked the \hbox{0.5--2~keV} and \hbox{2--8~keV} data of these
sources to obtain their average X-ray emission in each bandpass and infer their
mean spectral slopes.  The details of the data reduction and stacking procedure
used are described in detail in Section 4.2 of \citet{2008ApJ...681.1163L}.
The stacked X-ray spectrum was fitted with a power law model, giving the photon
index of $\Gamma = 1.53 \pm 0.09$. We further divided galaxies into two groups,
separated by the median X-ray flux, $f_{\rm{X}}\sim
10^{-16}\,\rm{erg}\,\rm{cm}^{-2}\,\rm{s}^{-1}$ and obtained the mean photon
indices of $\Gamma = 1.53 \pm 0.11$ and $\Gamma = 1.66 \pm 0.15$ for the bright
and faint groups respectively. For comparison, we stacked the sample of sources
detected in both X-ray and radio and obtain a mean photon index $\Gamma = 1.99
\pm 0.10$. 

The radio catalogs include only $\geq5\sigma$ detections. In order to probe deeper, we searched for 
sources on the radio maps down to $3\sigma$ and selected those within $1.5\arcsec$ from 
the positions of X-ray sources. We found ten sources, three in the CDF-N field and seven in the CDF-S field, 
one of which had not high enough specific SFR to pass our selection criterion (see Section 
\ref{sec:cdf_mstar}). We measured their radio intensities as follows. For the CDF-N we performed a Gaussian 
fit on both the $1.7\arcsec$ and $3\arcsec$ radio maps. Primary beam correction was applied along with a 
correction for radial bandwidth smearing. The results for each resolution image were then compared and the 
best fit was selected. If the peak SNR was more than 10\% higher on the lower-resolution image then that 
result was used \citep{2008AJ....136.1889O}. Finally, if the total flux density of the source was significantly 
larger at lower resolution then that result was adopted. In cases where the results appeared inconsistent, the 
images were examined to resolve the issue. One of the three CDF-N sources, $J123727.68+621036.3$ 
appears to be a possible FRI source \citep[see e.g.,][]{2003ApJS..146..267M}. It is extended in radio and 
therefore it was not included in the original radio catalog that we cross-matched (see Section\,
\ref{sec:cdf_sfr}). As this sources appears to be a radio-loud AGN with twin radio jets, we excluded it from the 
following analysis. For the ECDF-S field only mosaicked radio images are publicly available, 
therefore the radial bandwidth smearing correction could not be applied. We performed a Gaussian fit using 
the available maps and adopted the resulting peak flux for 
the radio counterparts to the X-ray sources. The X-ray stacking analysis of the sources with radio counterparts 
in the 3-5$\sigma$ range yielded an average $\Gamma = 1.30\pm0.17$, which is consistent with that 
obtained above for radio-undetected sources. We also investigated possible variations of the effective stacked 
photon index with the $L_{\rmn{X}}/\rmn{SFR}$ ratio and found no significant trend.

Thus, galaxies detected in both X-ray and radio bands have X-ray spectra
typical for star-forming galaxies \citep[e.g.][]{1999ApJS..120..179P},
confirming that their X-ray emission is indeed associated with star-formation.
On the other hand,  Chandra sources without radio counterparts  have harder
spectra,  $\Gamma \approx 1.4-1.6$, suggesting that there is likely to be some
level of low-luminosity and/or obscured AGN contamination within this
population.  Hereafter, we therefore exclude from the analysis of the
$L_{\rmn{X}}$--SFR relation X-ray sources undetected in the radio band. Sources with $3-5\sigma$ 
counterparts were also excluded.

\subsection{Stellar mass and specific star formation rate}
\label{sec:cdf_mstar}

As a final check, we verified that selected CDF galaxies satisfy the selection
criterion applied for the sample of resolved galaxies in Paper~I -- that their
specific SFR exceeds $\mathrm{SFR}/M_{\star} > 10^{-10}$ yr$^{-1}$. 

This was rather straightforward for late-type galaxies selected from
\citet{2008ApJ...681.1163L}, for which stellar mass measurements were
available.  For CDF-S galaxies from \citet{2011ApJS..195...10X}, we used the
multi-wavelength optical data provided in the same catalog. To be
consistent with the estimations in \citet{2008ApJ...681.1163L}, we calculated
the stellar masses using the calibration of \citet{2003ApJS..149..289B}.  In
this procedure, we made use of $(r-z)$ colors and the $K$-band luminosity.
One CDF-S source was
excluded because it did not have a $z_{850}-$band magnitude estimation. The
stellar mass was therefore obtained as follows:

\begin{equation}
\label{eq:mstar}
\log\bigg(\frac{M_{\star}}{M_{\odot}}\bigg) = \log\bigg(\frac{L_{\rmn{K}}}{L_{\rmn{K},\odot}}\bigg) - 0.092 + 0.019\,(r-z)
\end{equation}

In particular, we derived $L_{\rmn{K}}/L_{\rmn{K},\odot}$ based on K-band AB
magnitudes from the MUSYC catalog \citep{2009ApJS..183..295T}, and computed
$(r-z)$ colors based on AB magnitudes in R-band from the ESO 2.2 m WFI R-band
catalog \citep{2004ApJ...600L..93G}, and $z$-band magnitudes from the GOODS-S
Hubble Space Telescope (HST) catalog \citep{2004ApJ...600L..93G}. 

All CDF sources satisfied the  $\mathrm{SFR}/M_{\star} >1 \times 10^{-10}$ yr$^{-1}$ criterion.

\subsection{X-ray variability and AGN candidates in the CDF-S}
\label{sec:variability}

As an additional criterion to filter out AGN-contaminated galaxies, we used the results from \citet{2012ApJ...748..124Y} for the 4 Ms CDF-S, to separate star-forming galaxies from AGN candidates, based on their X-ray variability. We applied same criteria as in \citet{2012ApJ...752...46L}, namely: (i) X-ray variability on timescales of months to years detected  with the confidence of  $P_{\chi^2} \gtrsim 95\%$; (ii)  the 0.5--8 keV luminosity greater than $10^{41}\,\rm{erg}\,\rm{s}^{-1}$ \citep[see][for details]{2012ApJ...748..124Y}. By cross-matching our CDF-S sources with this catalog, we found that two sources, $J033206.42-274728.7$ and $J033218.04-274718.8$, are likely to be AGNs. These sources were excluded from the analysis.

\subsection{The final CDF sample}
\label{sec:CDF_sample}

The final CDF sample consists of 39 X-ray selected galaxies with radio SFR
measurements from the CDFs: 23 objects from CDF-N and
14 from CDF-S. They are listed in Table \ref{table:cdf_galaxies}, along
with their parameters. 

The entire sample, including 21 local (resolved) galaxies, 8 LIRGs and ULIRGs
and 37 CDF galaxies (66 galaxies in total) is shown in the SFR--$M_{\star}$
plane in Fig.\ref{fig:m-sfr}.

\section{{\boldmath $L_{\rmn{X}}-\rmn{SFR}$} relation for total luminosity}
\label{sec:lx_sfr_relation}

\begin{figure}
\centering
{
\includegraphics[trim=1mm 15mm 5mm 5mm, width=1.0\linewidth]{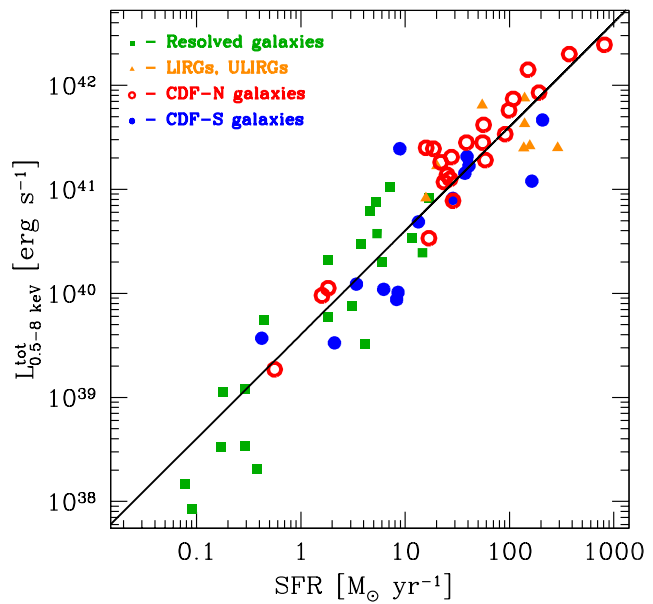}
}
\caption{The $L_{\rmn{X}}^{\rmn{tot}}-\rmn{SFR}$ relation. The (green) squares
indicate galaxies from the local ($z=0$) resolved sample, the (orange)
triangles are LIRGs and ULIRGs from the unresolved sample, described in Sect.
\ref{sec:resolved_sample} and presented in Table \ref{table:local_sample}. The
filled (blue) circles are late-type galaxies from the CDF-S
and the and empty (red) circles are late-type galaxies from the 
CDF-N, selected as described in Sect. \ref{sec:cdf} and presented in
Table \ref{table:cdf_galaxies}.  The solid line shows the linear scaling
relation, obtained using the whole sample of 54 galaxies and given by eq.
(\ref{eq:ltot_sfr_all}).}  
\label{fig:ltot_sfr}
\end{figure}

We approximate the $L_{\rmn{X}}^{\rmn{tot}}-\rmn{SFR}$ data (in
Fig.\ref{fig:ltot_sfr}) with  the log-linear model $\log
L^{\rmn{tot}}_{\rmn{X}} = \log K + \beta\,\log \rmn{SFR}$.  This model was
applied to local sample and CDF sample separately. Both slope $\beta$ and
normalization $K$ were set as free parameters of the fit. To fit the data we
used least square minimization. The results are given in Table
\ref{table:least_squares}, which demonstrates that the best fit parameters for
the local and CDF samples are compatible. We therefore performed the joint fit
to the whole sample of star-forming galaxies that includes both local and CDF
sources, using same model. We obtained a slope $\beta = 1.00\pm 0.05$,
suggesting a linear relation. The linear scaling relation obtained by fixing
the slope to unity is shown in Fig. \ref{fig:ltot_sfr} and it is given by the
following equation:
\begin{equation}
\label{eq:ltot_sfr_all}
L^{\rmn{tot}}_{0.5-8\,\rmn{keV}} (\rmn{erg}\,\rmn{s}^{-1}) \approx (4.0\pm 0.4)\times 10^{39}\,\rmn{SFR}\,(M_{\odot}\,\rmn{yr}^{-1})
\end{equation}
where the uncertainty was computed from the scatter of the points around the best fit. 
The results of all fits are summarized in Table \ref{table:least_squares}.

In Paper~I we obtained the  $L_X-SFR$  relation for X-ray point sources in
star-forming galaxies $\approx 2.6 \times 10^{39}$
($\rmn{erg}\,\rmn{s}^{-1})/(M_{\odot}\,\rmn{yr}^{-1}$). Comparing the scale
factor for compact sources with that in eq.(\ref{eq:ltot_sfr_all}) above we
conclude that the unresolved emission contributes, on average,  $\sim 1/3$ of
the total \hbox{0.5--8~keV} luminosity of star-forming galaxies generated per unit of SFR.
This result is consistent with the preliminary estimate obtained in Paper~I
based on the comparison of the $L_{\rmn{X}}-\rmn{SFR}$ relations of the HMXBs
in the resolved galaxies and total emission from unresolved galaxies. It is
also consistent with the scaling relation for the ISM emission obtained in
Paper~II.

In the conclusion of this section we note that the result of the fitting of the
$L_{\rmn{X}}-\rmn{SFR}$ relation depends on whether the fit  was performed in
the logarithmic  or linear space. In the linear case, it also depends on
whether the least square fit in the form $L_{\rmn{X}}=A\times \rmn{SFR}$ was
performed or individual $L_{\rmn{X}}/\rmn{SFR}$ ratios were averaged (note that
in the log case it is same). Note also that    large scatter of points, much
bigger than statistical uncertainties, renders the $\chi^2$ minimization less
appropriate. Doing the least square fit to the linear quantities, we obtain the
scale of $L_{\rmn{X}}/\rmn{SFR}\approx (2.7\pm 0.2)\times 10^{39}$
($\rmn{erg}\,\rmn{s}^{-1})/(M_{\odot}\,\rmn{yr}^{-1}$). Averaging the
$L_{\rmn{X}}/\rmn{SFR}$ ratio in the linear space, we obtain yet another
number: $L_{\rmn{X}}/\rmn{SFR}\approx (5.4\pm 0.6)\times 10^{39}$
($\rmn{erg}\,\rmn{s}^{-1})/(M_{\odot}\,\rmn{yr}^{-1}$), i.e. notably larger
than in eq.(\ref{eq:ltot_sfr_all}). The value obtained in the logarithmic  fit
is smaller, because it gives equal weights to the low and high
points.\footnote{Indeed, for two measurements yielding $y_1=a$ and $y_2=10\,a$
the logarithmic (geometrical) average is $\left<y\right>_{log}\approx 3.16\,a$
while the linear average is $\left<y\right>_{lin}= 5.5\,a$, i.e. is nearly
twice larger.} As this seems to be a more correct approach and in order to be
consistent with Paper~I, we used the fit in the log space as default.

\begin{table*}
\centering
\begin{minipage}{105mm}
\caption{Summary of the parameters for $L_{\rmn{X}}-\rmn{SFR}$ relations obtained from least-squares fit.}
\label{table:least_squares}
\begin{tabular}{@{}l l c c l c @{}}
\hline
\vspace{1mm}
&  \multicolumn{3}{|c|}{{\sc free slope}} &  \multicolumn{2}{|c|}{{\sc linear fit}}\\
Sample &\vline\, $\log K$ & $\beta$ & $\sigma$ & \vline\, $\log K$ & $\sigma$\\
 &\vline\, & & (dex) & \vline\, & (dex)\\
\hline
\hline
$z=0$ galaxies	 &\vline\, $39.53 \pm 0.09$ & $1.04 \pm 0.08$ & $0.42$ &\vline\, $39.56 \pm 0.08$ & $0.42$\\
CDF galaxies	&\vline\, $39.66 \pm 0.12$ & $ 0.98\pm 0.07$ & $0.33$ &\vline\, $39.64 \pm 0.05$ & $0.32$\\
CDF-S galaxies &\vline\,  $39.65\pm0.21$ & $0.85\pm0.16$ & $0.41$ &\vline\, $ 39.47\pm 0.11$ & $0.41$\\ 
CDF-N galaxies &\vline\,  $39.76 \pm 0.11 $ & $0.98\pm 0.06$ & $0.22$  &\vline\, $39.74 \pm 0.05$ & $0.21$\\ 
all &\vline\, $39.57 \pm 0.07$ & $ 1.03 \pm 0.05$ & $0.37$ &\vline\, $39.60\pm 0.05$  & $0.37$ \\
\hline
\end{tabular}
Note. The parameters are relative to the least-squares fit to the data with the
relation $\log L_{\rmn{X}} = \log K + \beta\,\log \rmn{SFR}$, respectively
setting the slope $\beta$ free and fixing it to unity. $\sigma$ is the
dispersion around the best-fitting relation. See Sect.
\ref{sec:nearby_Xray_lums} and \ref{sec:cdf_Xray_lums} for the definition of
the X-ray luminosities, Sect. 6 of Paper~I and \ref{sec:cdf_sfr} of present
work for the definition of SFR for $z=0$ sample and CDF galaxies respectively.
\medskip
\end{minipage}
\end{table*}

\subsection{CDF-S vs CDF-N}

Visual inspection of Fig.~\ref{fig:ltot_sfr} suggests that there may be
an offset in $L_{\rmn{X}}/\rmn{SFR}$ ratio between the CDF-S and CDF-N fields.
To asses its statistical significance, we fitted the data for the CDF-S and
CDF-N separately; the results are presented in Table \ref{table:least_squares}.
For the linear model, we found that the $L_{\rmn{X}}/\rmn{SFR}$ ratios for CDF-N and CDF-S
differ by 0.27 dex. The statistical significance of this difference measured
with respect to the dispersion of the points is $\approx 2.2\sigma$, i.e. the
offset is marginally significant. We also note that the CDF-N galaxies appear
to have smaller scatter, $\approx 0.2$ dex vs $\approx 0.4$ dex. 

To explore the origin of the difference between the fields,  we first
investigate importance of the   Eddington bias in our CDF samples. To this end,
we compared the fluxes of CDF-N and CDF-S sources with the sensitivity limits,
in the 0.5--8 keV band, of the respective fields, $7.1\times
10^{-17}$~erg~cm$^{-2}$~s$^{-1}$ for the CDF-N
\citep{2003AJ....126..539A} and $3.2\times
10^{-17}$~erg~cm$^{-2}$~s$^{-1}$ for the CDF-N
\citep{2011ApJS..195...10X}. We selected only CDF galaxies with flux above
three times the sensitivity limit of the given field. Such a selection shrank
the sample by $\sim$50\%. For the new sample we repeated the fitting procedure
and obtained $\log K = 39.56 \pm 0.15$ for CDF-S and $\log K = 39.59 \pm 0.08$
for CDF-N. These numbers are in perfect agreement with each other as well as
with the local sample (Table \ref{table:least_squares}) and, taken at the face
value, may suggest that the offset between CDF-S and CDF-N is caused by the
Eddington bias. We note however, that whereas for the CDF-N the $\log K$ shifts
downwards for the reduced sample, as it should be expected if X-ray fluxes were
affected by the Eddington bias,  CDF-S data shows the opposite behavior, with
the best fit $\log K$ increasing for the reduced sample.  Thus, change of the
$\log K$ in CDF-S and good agreement between the two fields in the reduced
sample was a result of statistical fluctuations.  We conclude that although the
Eddington bias may explain a part of the offset between CDF-S and CDF-N, it is
unlikely that it is responsible for all of it.

We can exclude that systematic differences in Chandra calibration of the CDF-N
and CDF-S would affect the observed $L_{\rmn{X}}/\rmn{SFR}$ offset. Indeed, the
1 Ms CDF-S and 2 Ms CDF-N were analyzed in exactly the same way by
\citet{2003AJ....126..539A}. The data for both surveys were taken throughout
the first three years of the mission, therefore comparing the 1 Ms CDF-S with 4
Ms CDF-S systematics would provide us with an approximation of the difference
in calibration between the CDF-N and CDF-S. In particular, sources with flux
$F_{0.5-8\,\rm{keV}} > 10^{-15}$~erg~cm$^{-2}$~s$^{-1}$ have mean
$F_{0.5-8\,\rm{keV}}(4\,\rm{Ms})/F_{0.5-8\,\rm{keV}}(1\,\rm{Ms}) = 0.994$, i.e.
a difference of $\sim0.5\%$, which is negligible for the aims of the present
work.  Furthermore, the amplitude of the difference $\Delta \log(K)\approx
0.27$ dex, corresponds to the factor of $\approx 1.9$ difference in flux
calibration, which is far beyond the calibration uncertainties expected for
Chandra.  

Another potential source of systematic difference in the $L_X/SFR$  between the two fields is offset in the radio data. This however requires a special investigation which is beyond the scope of this paper.

\subsection{Comparison with previous results}

\citet{2003A&A...399...39R} studied the $L_{\rmn{X}}-\rmn{SFR}$ relation using
X-ray data from ASCA and BeppoSAX satellites. In Paper~I, we converted their
X-ray luminosity and SFR to make them consistent with the analogous quantities
used here and derived: $L_{0.5-8\,\rmn{keV}}\approx 1.6\cdot 10^{39}\times
\rmn{SFR}$, i.e. more than $\sim 2$ times smaller than our
eq.(\ref{eq:ltot_sfr_all}). The rather large discrepancy is likely caused by
the confusion in the definition of SFR proxies (see detailed discussion in
Paper~I). On the other hand, the $L_{\rmn{X}}-L_{\rmn{IR}}$ relation  of
\citet{2003A&A...399...39R}, transformed  to be compatible with our definitions
is $L_{0.5-8\,\rmn{keV}}\approx 2.4\cdot 10^{-4}\times L_\rmn{IR}$. The scale
in this relation is $\approx 27\%$ larger that the scale in the corresponding
relation for HMXBs from Paper~I (eq.(23)), consistent with $\sim 1/4$
contribution of hot ISM and other unresolved emission components  to the total
luminosity   -- see the discussion after eq.(\ref{eq:ltot_sfr_all}) above. This
proves that the underlying $L_{\rmn{X}}-L_{\rmn{IR}}$ relations are fully
compatible and the discrepancy in the $L_{\rmn{X}}-\rmn{SFR}$ relation is due
the difference in the use and definitions of SFR proxies.

\citet{2007A&A...463..481P} investigated the relation between the total X-ray
emission from normal and starburst (ULIRGs) galaxies in the 2--10 keV band and
the SFR derived from infrared luminosity. Using their eq. (10) and converting
the 2--10 keV luminosity to the 0.5--8 keV band (a factor of 1.28) we obtain:
$L_{0.5-8\,\rmn{keV}}\approx 4.9 \cdot 10^{39}\times \rmn{SFR}$. This relation
is consistent, within a factor of $\sim 1.4$ with our  eq.
(\ref{eq:ltot_sfr_all}).  On the other hand, in Paper I we found a rather large
disagreement between our $L_{\rmn{X}}-\rmn{SFR}$ relations for the HMXB
luminosity.

\citet{2010ApJ...724..559L} recently studied the relation between 2--8 keV
luminosity and SFR for a sample of nearby LIRGs using {\it Chandra}
observations. They obtained the hard band luminosities by modeling the X-ray
spectra in 0.5--8 keV band using Galactic absorption, a thermal component and a
power-law component. This is consistent with the model we used to fit the X-ray
spectra of our sample of LIRGs and ULIRGs and determine their X-ray
luminosities. In order to compare our results with the $L_{\rmn{X}}-\rmn{SFR}$
relation obtained by \citet{2010ApJ...724..559L}, we converted their 2--8 keV
luminosity to the 0.5--8 keV band. We assumed same model as above, with average
quantities from our ULIRGs spectral fits ($\Gamma \sim 1.7$, $kT \sim 0.6$ keV,
$N_{\rmn{H}}\sim 2\times 10^{21}\,\rmn{cm}^{-2}$ and thermal-to-power-law
normalization ratio of $\sim 0.5$). For the linear (first line in their Table
4) $L_\rmn{X}-\rmn{SFR}$ relation in \citet{2010ApJ...724..559L} we obtain a
scale factor $L_\rmn{X}/\rmn{SFR} \approx 2.9\times10^{39}$
($\rmn{erg}\,\rmn{s}^{-1})/(M_{\odot}\,\rmn{yr}^{-1}$). This result  is also
in a reasonable agreement with our eq. (\ref{eq:ltot_sfr_all}).

\section{Constraints on the redshift and SFR dependence of the $L_{\rmn{X}}-\rmn{SFR}$ relation}

\begin{figure*}
\centering
\hbox{
\includegraphics[width=0.5\linewidth]{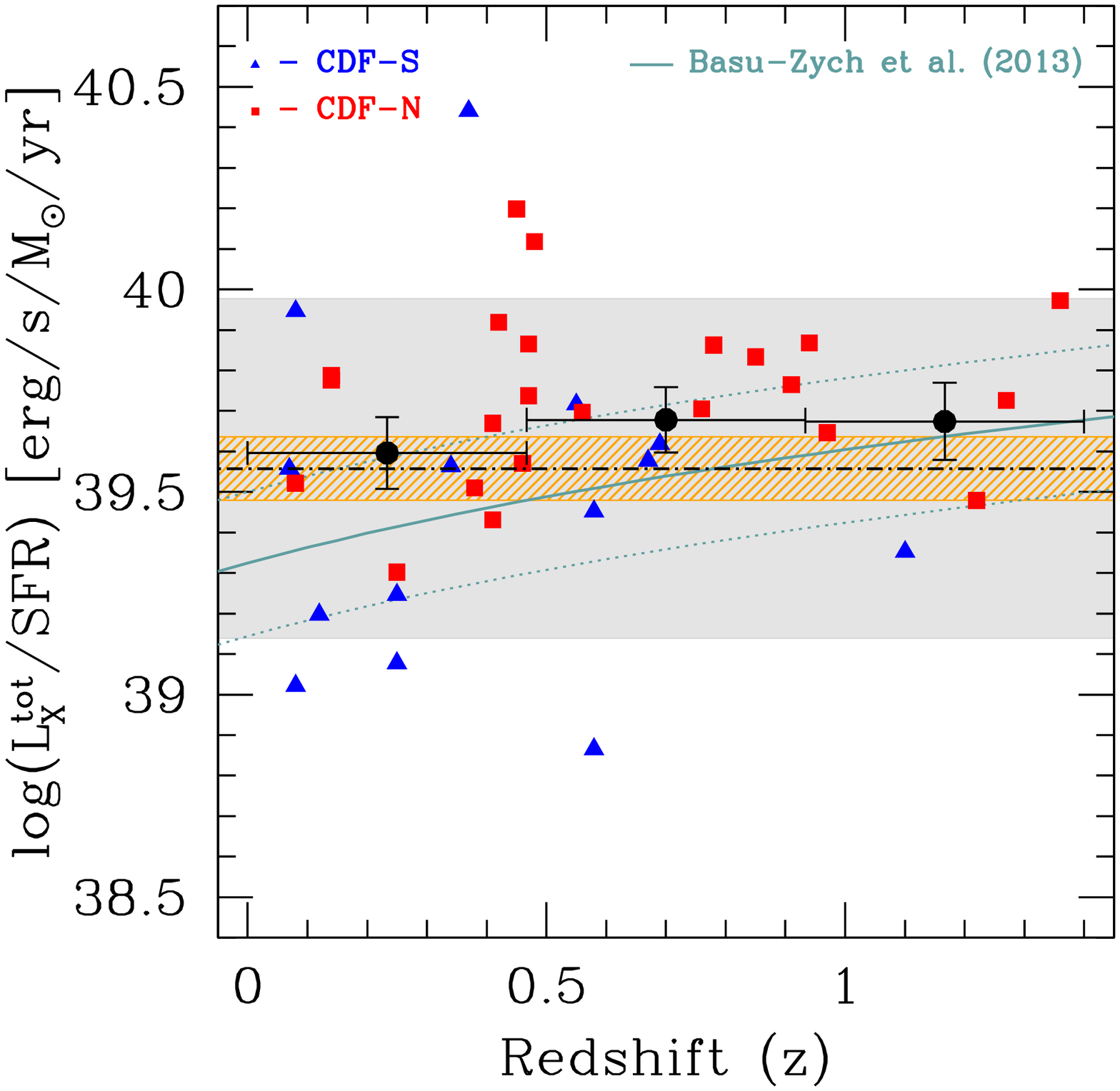}
\includegraphics[width=0.5\linewidth]{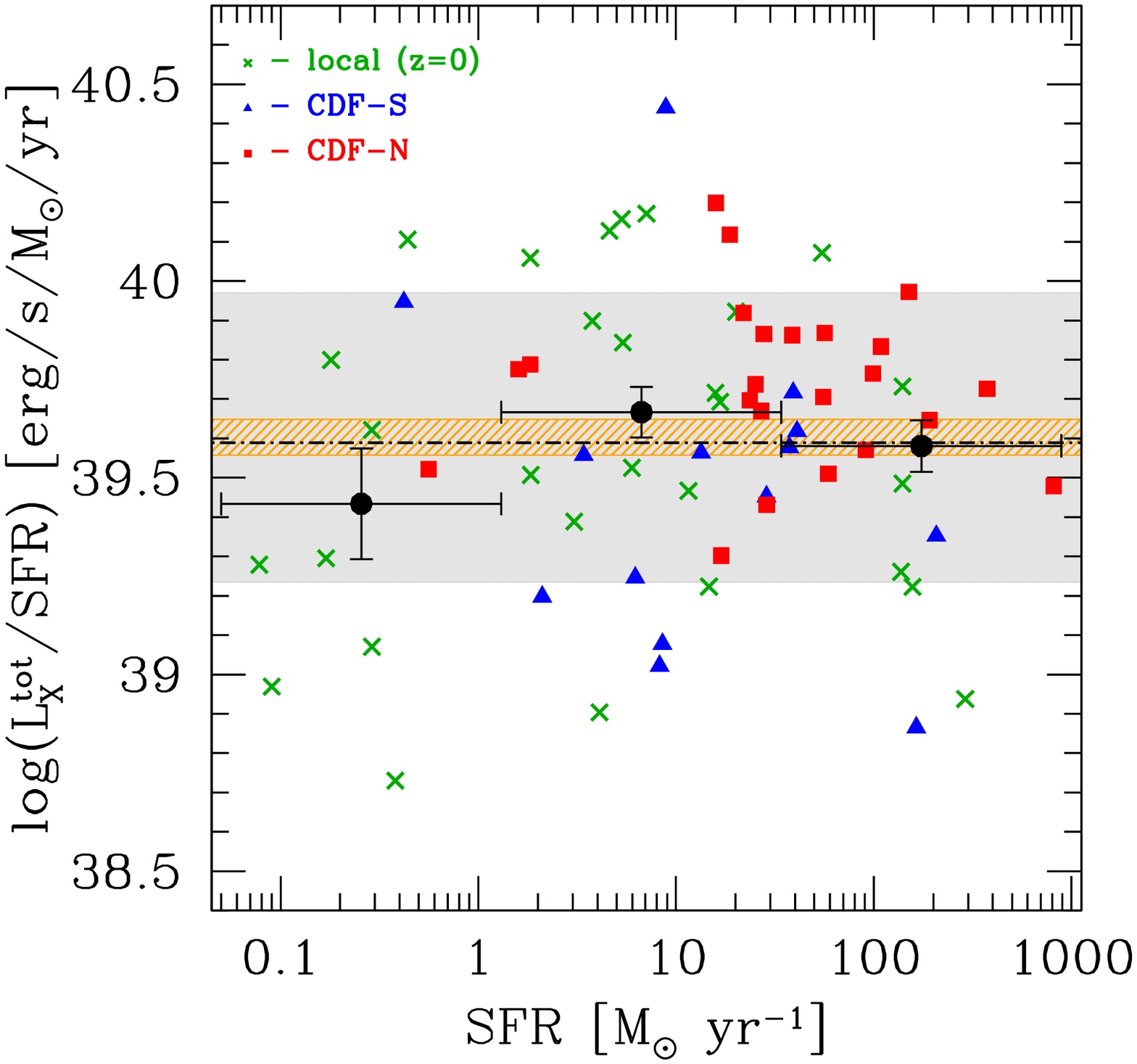}
}
\caption{Dependence of the $L_{\rmn{X}}^{\rmn{tot}}-\rmn{SFR}$ relation on
redshift (left panel) and star formation rate (right panel).The symbols without
error bar indicate individual galaxies. The (black) filled circles with error
bars show their average values within each redshift (left) or SFR (right) bin.
The vertical error bars are uncertainties of the average values computed from
the {\em rms} of points in the given bin. In the left panel, the horizontal
dot-dashed line shows the average local  $L_{\rmn{X}}^{\rmn{tot}}/\rmn{SFR}$
ratio.  Its uncertainty is shown by the narrower (orange) shaded area, while
the wider (grey) shaded area shows the $rms$ of the points. On the right hand
panel, the horizontal line and the shaded areas show similar quantities
computed using all galaxies. For comparison, in the left-hand panel we
plot the best-fit model from \citet{2013ApJ...762...45B} stacking results that
is appropriate for the median SFR (solid, grey curve) and interquartile
SFR range (dotted, grey curves) of our CDF galaxy sample.
}
\label{fig:z_dep}
\end{figure*}

The composition of the sample (29 galaxies at the redshift $z=0$ and 25
galaxies at $z\approx 0.1-1.3$) allows us to constrain a possible redshift
evolution of the X-ray emission produced per unit SFR. To this end, we grouped
the CDF galaxies into three redshift bins of equal width spanning from $z=0$ to
$z=1.3$ and computed the average $L_{\rmn{X}}^{\rmn{tot}}/\rmn{SFR}$ ratio and
its uncertainty within each bin.  Same quantities were computed for the local
sample of galaxies, to characterize the $z=0$ population. The result of this
calculation is shown in Fig. \ref{fig:z_dep}.  It is evident from the figure,
that the average values for CDF galaxies are consistent, within uncertainties,
between each other and with the average quantity for local galaxies.
To confirm this, we applied the Spearman's rank test and we found no
statistically significant correlations, neither for the individual values nor
for the average points, with the corresponding probabilities of $P\sim 27\%$
and $P\sim 20\%$  respectively.

Thus,  the total X-ray emission produced per unit SFR by late-type galaxies
does not show statistically significant trends with the redshift up to $z\sim
1.3$. The sensitivity of this analysis is sufficient to rule-out evolution
larger than $\sim 0.1-0.2$ dex (by factor of $\sim 1.3-1.6$) between $z \sim 0$ and $z \sim 1.3$, although  as a
caveat  we should note that the high redshift bin contained only 4 galaxies.
This conclusion  is in agreement with results of the  X-ray stacking analysis
of late-type galaxies \citep{2008ApJ...681.1163L} and with calculations of the
maximal contribution of star-forming galaxies to the unresolved CXB intensity
\citep{2012MNRAS.421..213D}. 

Recent population synthesis calculations by \citet{2012arXiv1206.2395F} predict
that the ratio of the 0.5--8 keV luminosity to the  SFR increases by a factor
of $\sim 1.3 - 1.4$ between redshift $z=0$ and $z\sim 1.3$.  Such an evolution
is within the uncertainties of our results. 

Furthermore, recent deep X-ray stacking of the 4 Ms CDF-S data,
\citet{2013ApJ...762...45B} suggested that there may be mild evolution of the
rest-frame 2--10 keV luminosity per unit SFR with the redshift and SFR.   In
Figure~\ref{fig:z_dep}, we show the expected redshift evolution of $L_{\rm
X}$/SFR as a function of redshift, from \citet{2013ApJ...762...45B}, for SFRs
equal to the median SFR ({\it solid grey curve\/}) and interquartile SFR range
({\it dotted grey curve\/}) of our CDF galaxy sample.  These curves have been
corrected to our choice of X-ray bandpass and IMF.  It is obvious from
Figure~\ref{fig:z_dep} that although our data does not require redshift
evolution of the scale in the mean $L_{\rm X}$/SFR relation, it is also
consistent with the mild evolution found by \citet{2013ApJ...762...45B}.  

Finally, in the right hand panel of Fig.\ref{fig:z_dep} we constrain the
possible dependence of the $L_{\rmn{X}}/\rmn{SFR}$ ratio on the SFR. Using both
local and CDF galaxies, we bin the data into 3 SFR bins and compute the average
and its uncertainty ($rms$-based) for each redshift bin.  The formal
$\chi^2=4.8$ for the binned data, with 2 degrees of freedom.  The
probability of having such (or bigger) value of the $\chi^2$ is $\sim 9\%$,
which is equivalent to $\sim 1.5\,\sigma$ detection.  This may  or may not
indicate the existence of some SFR dependence in the $L_{\rmn{X}}-\rmn{SFR}$
ratio. If any, the amplitude of this dependence also does not exceed  $\sim
0.15$ dex (a factor of $\sim 1.4$). Finally, we note that due
to the observer's bias (see Sect. 2.3 of Paper~I), the effect of
statistics of the small numbers \citep{2004MNRAS.351.1365G} is not strong in
our sample.

\section{Summary}

Based on the sample of nearby resolved galaxies more distant ULIRGs at
intermediate distances and star-forming galaxies from the CDFs, we construct a
sample of 54 star-forming galaxies spanning the range of redshifts from $z=0$
up to $z=1.3$ and the range of star formation rates $\rmn{SFR}\sim 0.1-10^3$
M$_\odot$/yr (Fig.\ref{fig:m-sfr}). Using this sample, we calibrate the
$L_{\rmn{X}}-\rmn{SFR}$ relation for the 0.5--8 keV band luminosity
(Fig.\ref{fig:ltot_sfr}). We find that $L_{\rmn{X}}-\rmn{SFR}$ dependences for
the  local and CDF samples are consistent with linear relations with the
typical accuracy of $\sim 0.1$ in the slope. The linear
$L_{\rmn{X}}-\rmn{SFR}$ relation obtained for the entire sample is given by
eq.(\ref{eq:ltot_sfr_all}). We did not find any statistically significant
trends in the scaling relation  with the redshift and star formation rate with
the upper limit on the possible variations in the  $L_{\rmn{X}}/\rmn{SFR}$
ratio of $\sim 0.1-0.2$ dex (a factor of $\sim 1.3-2.6$) (Fig.\ref{fig:z_dep}).
This property makes the X-ray emission a powerful tool to measure star
formation rate in a broad range of redshifts and star formation regimes, which
can be applied {\em en masse} to faint distant galaxies.

\section*{Acknowledgments}

SM gratefully acknowledges financial support through the NASA grant AR1-12008X and funding from the 
STFC grant 664 ST/K000861/1.The authors are grateful to Ken Kellermann for providing them with both
radio RMS map and the source list of the second data release of the VLA
$1.4\,\rm{GHz}$ Survey of the Extended {\it Chandra} Deep Field South, prior to publication. 
The authors thank William Forman for his valuable comments and suggestions to improve the
quality of the paper.  This research made use of \textit{Chandra} archival data
and software provided by the \textit{Chandra} X-ray Center (CXC) in the
application package CIAO. This research has made use of SAOImage DS9, developed
by Smithsonian Astrophysical Observatory.  The \textit{Spitzer Space Telescope}
is operated by the Jet Propulsion Laboratory, California Institute of
Technology, under contract with the NASA.  \textit{GALEX} is a NASA Small
Explorer, launched in 2003 April.  This publication makes use of data products
from Two Micron All Sky Survey, which is a joint project of the University of
Massachusetts and the Infrared Processing and Analysis Center/California
Institute of Technology, funded by the NASA and the National Science
Foundation.  This research has made use of the NASA/IPAC Extragalactic Database
(NED) which is operated by the Jet Propulsion Laboratory, California Institute
of Technology, under contract with the National Aeronautics and Space
Administration.

\bsp

\label{lastpage}

\end{document}